\documentclass{frontiersFPHY} 

\usepackage{url,lineno}

\copyrightyear{}
\pubyear{}

\def\journal{Stellar and Solar Physics}
\def\DOI{}
\def\articleType{Specialty Grand Challenge}

\def\firstAuthorLast{M.J. Thompson} 
\def\Authors{Michael J. Thompson\,$^{1,*}$}
\def\Address{$^{1}$High Altitude Observatory, National Center for Atmospheric Research, Boulder, CO, USA }
\def\corrAuthor{Michael J. Thompson}
\def\corrAddress{National Center for Atmospheric Research, P.O. Box 3000, Boulder, CO 80307-3000, USA}
\def\corrEmail{mjt@ucar.edu}

\begin{document}
\onecolumn
\firstpage{1}

\title[Grand challenges in solar physics]{Grand Challenges in the Physics of the Sun and Sun-like Stars}
\author[\firstAuthorLast ]{\Authors}
\address{}
\correspondance{}
\extraAuth{}
\topic{}

\maketitle


%
%

\section{Introduction}

``If the Sun had no magnetic field, it would be as 
uninteresting as most astronomers think it is''. This statement 
is attributed to 
R.~B.~Leighton \citep{Moore1985}. Personally I think the Sun is of enormous
interest in all respects, magnetic and non-magnetic; 
nonetheless, the grand challenges 
I have selected for this article 
do indeed pertain to the Sun's magnetic field. 

The study of stellar structure and evolution is one of the main building 
blocks of astrophysics, and the Sun has an importance 
both as the star that is most amenable to detailed study and as the star that 
has by far the biggest impact on the Earth and near-Earth environment 
through its radiative and particulate outputs. 
Over the past decades, studies of stars and of the Sun have 
become somewhat separate. But in recent years, the rapid advances in 
asteroseismology, as well as the quest to better understand solar and 
stellar dynamos, have emphasized once again the synergy between 
studies of the stars and the Sun. In this article I have selected two ``grand
challenges'' both for their crucial importance and because I thnk that these
two problems are tractable to significant progress in the next decade. 
They are 
(i) understanding how solar and stellar dynamos generate magnetic field,
and (ii) improving the predictability of geo-effective space weather.

\section{Solar and Stellar Dynamos}

{\em How does the Sun generate its periodically reversing large-scale magnetic 
field? How do other solar-like stars generate their magnetic fields, and 
what are the similarities and differences between stellar activity cycles 
and that of the Sun? What can be learned about the solar dynamo 
by studying other stars?}

One of the most evident manifestations of solar magnetism is the number of 
sunspots, which waxes and wanes with an approximately 11-year quasi-periodic 
cycle. Once the polarity flip between 11-year cycles is taken into account, 
this becomes an approximately 22-year cycle. The Sun's large-scale ambient 
field, which is predominantly dipolar, has a similar 22-year cycle. Sunspots occur where concentrations of magnetic 
flux poke out through the Sun's surface, inhibiting the convection and 
causing that portion of the surface to be cooler (and hence darker) 
than its surroundings. Sunspots often occur in identifiable bipolar pairs, roughly oriented along lines of constant latitude but with the leading spot 
typically closer to the equator. The polarity of the leading spot is 
oppositely signed in the two 
hemispheres, and moreover changes sign every approximately 11 years.

The number of sunspots reaches a maximum approximately every eleven years, 
though the cycle length is quite variable. Also, the number of sunspots at 
maximum is very variable. There can also be extended periods when the sunspot
cycle appears to turn off, notably in the Maunder minimum of approximately
1645-1715, and proxies for solar activity such as isotope deposits in ice
cores suggest that such ``grand minima'' occur occasionally and apparently 
randomly in the Sun's past. 

How the Sun generates its oscillatory magnetic field is, 
however, not yet understood. The appearance of pairs of sunspots of opposite
polarity is strongly suggestive of magnetic flux tubes rising from the 
interior and that these tubes are approximately aligned parallel to the 
solar equator, i.e., the field they contain is {\it toroidal}. The large-scale 
but weaker field is {\it poloidal}. It seems likely that in some way the
cycling large-scale magnetic field of the Sun involves a dynamo in the
course of whose operation toroidal field is converted to poloidal field and 
poloidal field is converted to toroidal field. Typically, generic dynamo 
models involve stretching, twisting and folding of the magnetic field
\citep[e.g.][]{Childress1995}.

There are a number of models for how the Sun continues to generate a 
large-scale magnetic field via dynamo action, but none at this point
is anything more than a cartoon of what may be taking place. Many are 
``mean-field dynamos'', which are based on the assumption that one can
make a separation of scales between the large scale on which one wishes to 
describe the evolution of magnetic field, and the small-scale interactions 
of magnetic field and plasma motions that ultimately get parametrized in 
some closure scheme \citep[e.g.][]{Moffatt1978}.
Some often-invoked elements for the solar dynamo
are differential rotation in the solar interior, which stretches out 
the poloidal field to produce toroidal field, and the convective motions
that take place in the convective envelope of the Sun, which may take
toroidal field and produce poloidal field via the so-called 
{\em alpha} mechanism.  Helioseismology has 
mapped the rotation in much of the solar interior
\citep[e.g.][]{Thompson1996, Schou1998}. Helioseismology has also established
that the convective envelope occupies the outer 30 per cent of the solar 
interior \citep{jcd1991}, and that the base of the convection zone roughly coincides with 
a region of rotational shear that is now called the tachocline
\citep[e.g.][]{jcd2007}. 

The state of understanding of the solar dynamo has been reviewed by e.g. 
\cite{Weiss2009}, \cite{Charbonneau2010} and 
\cite{Jones2010}. A class of models that is currently 
popular in solar physics is the ``flux transport dynamo model'' 
\citep{Choudhuri:etal:1995, Dikpati1999}. These models are mean-field models 
that invoke the Babcock-Leighton mechanism in which the near-surface motions 
of differential 
rotation and meridional circulation statistically convert the toroidal field 
in decaying sunspots into a poloidal field. Meridional circulation sets up 
a conveyor belt that advects this field to high latitudes, 
then subducts it to the base of the convection and transports it towards 
the equator, during which passage it gets converted to toroidal flux that
rises to the surface when it gets strong enough to be magnetically buoyant 
and forms sunspots.

Shortcomings of present-day models of the solar dynamo are that they
are either highly idealized mathematical or computational models that possibly 
elucidate some of the principles but do not yet match the solar behavior; or 
they have {\it ad hoc} parameters that can match the observed 
large-scale behavior (e.g. sunspot number) but have little or no
predictive power. That we are still far from a robust predictability is well
illustrated by the wide range of predictions for the amplitude of Cycle 24
\citep{Pesnell2008},
most of which inevitably were incorrect.

Recent developments include numerical models of the solar convection zone and
outer radiative interior that capture the convective motions and rotation
and begin to show cycling dynamo behaviour 
\citep{Brown2010, Brown2011, Ghizarou2010, Racine2011}, 
though they do not yet succeed in
producing solar-like behavior: either they need a rotation rate that is 
far greater than that of the Sun, or they produce cycle periods that are longer 
than the Sun's. Nonetheless this line of research is promising. 
Understanding the solar dynamo is certainly a Grand Challenge.

Other stars are also observed to exhibit magnetic activity cycles
\citep[e.g.][and references therein]{Judge2012}, 
and seeking to 
understand stellar activity cycles and the Sun's dynamo in the context of 
those of other stars is a promising line of attack on the solar dynamo 
problem. Asteroseismology is opening up the study of stellar interiors, 
analogous to the impact of helioseismology on solar interior studies, and 
the {\it Kepler} mission in particular has made a step-change in the subject
\citep[e.g.][]{Chaplin2010, Chaplin2011, Chaplin2014, Metcalfe2010, Metcalfe2012, Beck2011}. For a summary of early and more recent asteroseismic results from
the {\it Kepler} mission, see, respectively, \cite{jcd2011} 
and \cite{Chaplin2013}.
For understanding stellar dynamos and the physical ingredients for 
dynamo action,
asteroseismology provides a valuable complement to traditional spectroscopy and
accuate photometry, which themselves are extremely useful for measuring 
stellar surface rotation rates and latitudinal differential rotation, as well as
revealing acitivity cycles similar to that of the Sun. A puzzle still to be 
resolved is that the Sun appears to be anomalous in the context of other
stellar dynamos. As shown by \cite{Bohm-Vitense2007}, activity cycle periods 
in a variety of other stars seem to fall onto two branches: those for which 
the cycle period $P_{\rm cycle}$ is about 400 times as long as the 
rotational period $P_{\rm rot}$ of the star, and those for which 
$P_{\rm cycle}$ is about 90 times as long as $P_{\rm rot}$. Some stars seem
to have two periods in their activity, one falling on each of two branches. 
This finding suggests there may be two basic dynamo modes in stars.
The Sun's 11-year cycle and approximatly 26-day rotation period puts it on
neither of these branches, but rather intermediate between the two. 
Interestingly, the Sun seems to exhibit a secondary period of about 2 years in
some of its activity indices, which would mean that the Sun's two activity 
periods are in a ratio that is not dissimilar to 400:90. There is much still 
to be understood.

\section{Improving the Predictability of Space Weather}

{\em What causes large potentially Earth-impacting space weather events on 
the Sun and how can we better predict them? What improvements, 
especially in terms of observations of the solar atmosphere and its magnetic 
field, can we foresee to improve forecasts of the geo-effectiveness of such 
events?}

As our nearest star, the Sun has a dominant influence on the Earth and 
near-Earth environment. One particular class of solar influences on the Earth 
is known collectively as space weather, magnetically driven episodic 
variations in the Sun's radiative and particulate outputs that impact on the 
Earth and geospace. The potential societal impacts of space weather -- on 
power grids, on communications and GPS, on satellites, on airline crew and passengers, on humans in space -- are increasingly recognized 
\citep{Baker2008}. 

The Sun's role as the driver of space weather is evident, but we have only 
a poor understanding of the physics that actually triggers the most impactful 
space weather eruptions -- X-class flares and coronal mass ejections (CMEs) -- and we have little capability to predict when such events will occur and 
how geo-effective they will be. To the latter point, it is particularly 
important to be able to determine whether the embedded magnetic field in an
Earthward-directed CME will be northward or southward, since the southward case 
is much more impactful as it interacts with the Earth's magnetosphere. 

Advances are needed in a number of key areas. New instrumentation and 
analysis tools are required to better observe the Sun's chromosphere
and corona and hence to determine the plasma conditions and magnetic fields
there. In contrast, the photosphere is relatively well observed and understood, 
though even there recent observations have thrown up surprises and controversy,
such as the finding from Hinode satellite observations that the small-scale
magnetic field is apparently predominantly horizontal rather than vertical
\citep{Lites2008}.

The Daniel K. Inouye Solar Telescope ({\it DKIST}), 
formerly known as the Advanced
Technology Solar Telescope ({\it ATST}), will be the largest ground-based solar 
telescope and will provide extremely high resolution observations of the Sun's
photosphere, chromosphere and corona, but only in a very small field of 
view \citep{Keil2003}. Though with its small field of view it will not provide 
a forecasting capability for space weather, a major justification for 
{\it DKIST} is to observe and 
lead to an understanding of the small-scale drivers of space weather events.
In my view, a key component for predicting the onset of large flares and 
possibly CMEs is also a knowledge of the near subsurface emergent magnetic 
field and plasma flows, and the only viable means of detecting these is 
local helioseismology \citep{Gizon2005}. There is evidence that the onset of 
major flare activity is preceded by an increase in kinetic helicity 
in the subsurface region \citep{Komm2009, Reinard2010}. Advances require
improved local helioseismic analyses, particularly in regions of strong
magnetic field \citep[e.g.][]{Braun2011}. Overall, a complete picture will
likely require better
theoretical understanding of the roles of a number of different elements --
magnetic field-line footpoint motions
in the photosphere, new emergent flux, and the complexity of existing 
magnetic fields in the solar atmosphere -- in
the genesis of space-weather events.

The chromosphere constitutes a boundary layer between the photosphere and solar 
interior on the one side and the corona and heliosphere on the other. 
It is the most poorly understood region of the solar atmosphere: it is 
highly dynamic in nature \citep[e.g.][]{DePontieu2007}, 
and the approximation of local thermodynamic equilibrium (LTE) is 
inadequate for modeling the observations there.
Yet it is a region through which mass 
and energy fluxes from the Sun must pass, and it can be argued that the
chromosphere rather than say the photosphere is the true bottom boundary 
for modeling the heliosphere and understanding space weather. 
\cite{Ayres2009}
provides a good overview of the challenges and opportunities for 
advancing understanding of the chromosphere and also gives context from
chromospheres of other stars. Spectro-polarimetric observations in 
multi-wavelengths of the spectral lines formed in the 
solar chromosphere are one
key to advancing understanding of this region, and a number of instruments
have been or are being developed and deployed to make such observations. These
include several of the first-light instruments to be deployed on 
the {\it DKIST}, the
{\it CRISP} instrument at the Swedish Solar Telescope \citep{Scharmer2008},
and the Chromosphere and Prominence Magnetometer, {\it ChroMag} 
\citep{DeWijn2012}.
Development of non-LTE spectro-polarimetric inversion codes 
\citep[e.g.][]{Socas-Navarro2000} is also essential for the interpretation
of the observations from the new suite of instruments.

It is evident from the spectacular loop structures observed there that the 
corona is dominated by magnetic fields, but direct measurements of the 
magnetic field in the corona are very challenging and are only now being 
realized \citep{Lin2000}.
Such measurements over extended spatial regions, 
complemented by magnetic field measurements in the chromosphere, promise to 
provide knowledge of the magnetic field in CMEs as they leave the Sun,
thus perhaps making possible the forecasting of the magnetic field strength
and direction in the sheaths and cores of Earth-impacting CMEs. 
Spectro-polarimetric observations in 
emission lines formed in the corona and 
observations at radio wavelengths provide two complementary avenues for
coronal magnetic field measurements. The proposed Frequency Agile Solar
Radiotelescope (FASR) will observe the corona at radio wavelengths
\citep{Bastian2003}. 
Observations in the near-infrared will be made by the 
{\it DKIST} (again, only in 
a small field of view) and by the proposed Coronal Solar Magnetism Observatory
coronagraph ({\it COSMO}) \citep{Gallagher2012}. 
A prototype for the {\it COSMO} coronagraph,
the Coronal Multi-Channel Polarimeter ({\it CoMP}), is currently making daily
spectro-polarimetric observations in the near-IR and has demonstrated that
it is possible to measure the magnetic field in the corona
\citep{Tomczyk2007, Tomczyk2008}. Modeling of observations to reconstruct the 
coronal magnetic field by tomographic or other techniques looks promising
\citep{Kramar2013}.

\section{Other issues}

I have chosen the above two areas of major challenge because of their importance
and because I believe that significant progress on 
them can be made in the next decade. No doubt, another author could have 
picked two other but equally fascinating areas of challenge. In closing, I 
would just like to mention a further set of issues that are undoubtedly 
important for improving our understanding of the Sun and Sun-like stars.

Since around 2005, there has been an ``abundance problem'' with the Sun. 
Prior to that date, solar models constructed with the then-current estimates
of the Sun's chemical abundances were in good agreement with helioseismology. But new spectroscopic analyses and 3-D atmospheric modeling 
by \cite{Asplund2005} revised significantly downwards the solar heavy-element 
abundance, particularly the oxygen abundance. This resulted in a much worse
agreement between solar models and the Sun's internal stratification as 
inferred from helioseismology \citep{Montalban2006}. Subsequent spectroscopic
re-evaluations of the solar abundances, though they have revised upwards 
slightly the values originally published by Asplund et al., still give a
significantly lower heavy-element abundance than pre-2005, and attempts to 
modify the microphysics assumed in 1-D stellar models have not
resolved the discrepancy with helioseismology \citep{Basu2013}. 
A number of current attempts
are underway in 2-D and 3-D models to incorporate macrophysics that
has not to date been part of the standard solar and stellar models. These
include incorporating rotation, magnetic fields, and internal gravity waves
\citep[e.g.][]{Talon2008, Eggenberger2010, Mathis2010, Mathis2013}.
These additional physical effects can variously 
redistribute angular momentum and chemical abundances within the stellar 
interior. Asteroseismology provides constraints on what can be assumed 
\citep{Deheuvels2012, Deheuvels2014}. My own view is that fully incorporating
these effects into models, particular in 3-D, may take rather longer than a 
decade. Nonetheless it is excellent that these modeling efforts have begun, and there will be a rich interplay between the modeling and asteroseismology for
years to come.

\section*{Acknowledgement}
I thank Scott McIntosh for comments that have improved this paper.
The National Center for Atmospheric Research is sponsored by the 
National Science Foundation.

\bibliographystyle{frontiersinHLTH&FPHY} 
\bibliography{Thompson-GC}

\end{document}